\documentclass[12pt]{iopart}
\usepackage{iopams}  
\usepackage{epsfig}
\usepackage{graphicx}
\usepackage{verbatim}

\begin{document}

\title[$H^+_2$ in a strong magnetic field]{$H^+_2$ in a strong magnetic field
described via  a solvable model}

\author{  R. D. Benguria\dag, R. Brummelhuis\ddag, P. Duclos\S, and S.
P\'erez-Oyarz\'un\P}

\address{\dag\ Facultad de F\'{\i}sica, Pontificia Universidad Cat\'olica de
Chile Casilla 306, Santiago 22, Chile, email: rbenguri@fis.puc.cl}

\address{\ddag\ School of Economics, Mathematics and Statistics, 7-15 Gresse
Street, 
University of London, U.K., e-mail: r.brummelhuis@statistics.bbk.ac.uk}

\address{\S\ Centre de Physique Th\'eorique - CNRS - Luminy
Case 907, F-13288 Marseille Cedex 9 and Phymat, Universit\'e de Toulon et du
Var,
83957 La Garde Cedex, France, email:  duclos@univ-tln.fr }

\address{\P\  Departamento de F\'{\i}sica, FCFM,  Universidad de Chile Casilla
487, Santiago 3, Chile, e-mail: sperez@puc.cl}

\begin{abstract}
We consider the hydrogen molecular ion $H^+_2$ in the presence of a strong
homogeneous
magnetic field. In this regime, the effective Hamiltonian is almost
one dimensional with a potential energy which looks like a sum of two Dirac delta
functions.
This model  is  solvable,  but not close enough to our exact Hamiltonian for relevant strenght
of the magnnetic field. However we show that the correct values of the equilibrium distance
as well as   the binding energy 
 of  the ground state of the ion, can be obtained when incorporating perturbative corrections up to
second order. Finally, we show that $ He_2^{3+}$ exists for sufficiently large
 magnetic fields.
\end{abstract}



\maketitle

\section{Introduction}

Since the discovery of strong magnetic fields in astronomical objects, e.g., in
the surface of neutron stars, 
the behaviour of atoms and molecules in these media has become a subject of wide
interest. Also, high-magnetic-field conditions can be mimicked in some semiconductors
where a small effective electron mass $m_*$ and a large dielectric constant $\epsilon$
reduce the Coulomb force relative to the magnetic force \cite{lai1}. For a review and references see
\cite{ruder,schmelcher,lai1,ruderman}.
In particular, the hydrogen molecular ion  $H_2^+$, has been studied under the
influence of a strong magnetic
field in the last 25 years, see e.g.,
\cite{bhaduri,lai,guillou,lopez,melo1,melo2,kashiev,
kravchenko,ozaki,peek,vincke,wunner,wille} for some references,
using variational and numerical techniques. This
system, as the simplest molecule in nature, can exhibit a qualitative behaviour
of the
internuclear separation as a function of the magnetic field that could have
applications to
the study of more complex molecular systems.


We consider  $H^+_2$, in a constant magnetic field. We use the Born-Oppenheimer
approximation, 
with $-e$ and $m_e$ the electric charge and the mass of the electron,
respectively. 
In Gaussian units the Hamiltonian is given by 
\begin{equation}
H = \frac{1}{2m_e}  \left( \vec p + \frac e c  \vec A \right)^2 - eV  +
\frac{e^2}{R} + H_{\rm{spin}},
\end{equation}
where
\begin{equation}
V= e\left(\frac{1}{|\vec r - \frac{R}{2} \hat z|} + \frac{1}{|\vec r +
\frac{R}{2} \hat z|}\right) 
\end{equation}
is the Coulomb potential due to the nuclei, and $R$  the internuclear
separation. Here 
$
H_{\rm{spin}} =  \frac{e}{m_ec} \vec S \cdot \vec B
$ 
and $\vec A $ is the magnetic vector potential.

In this static situation it  is convenient to use the Coulomb gauge $\vec
\nabla \cdot \vec A = 0$, and the choice 
$\vec A =  -\frac{1}{2} \vec B \times \vec r$. The magnetic field $\vec B$ is
taken parallel to  the unit vector 
$\hat z$ of the nuclear axis, which is the energetically most favorable
situation \cite{wille,guillou}.

The criterium to define the strong magnetic field regime is to ask for a 
gap between Landau levels bigger than an energy of the order of 
the ionization energy of the hydrogen atom. This means
\begin{equation}
\hbar w_0 = \frac{e^2}{a_0} =  27.2  \quad \!\!\!\! eV = 1   \quad \!\!\!\!  \textrm{Hartree},
\end{equation}
where $\omega_0 = eB_0/m_ec= 4.13\times 10^{16} \quad \textrm{seg}^{-1}$ 
is the cyclotron frequency and $a_0=\hbar^2/m_e e^2=0.53$ \AA, the Bohr radius.
That gives us the threshold value of $B$, 
\begin{equation}
B_0 = \frac{m_e^2 e^3 c}{\hbar^3} = 2.35 \times 10^9 \quad \textrm{G}.
\end{equation}
The Hamiltonian can be written as
\begin{equation}
H =  \hbar \omega_0 \left[ \frac{a_0^2}{\hbar^2} \frac{\vec{p}^2}{2} -
\frac{1}{2} 
\left(\frac{B}{B_0}\right) \frac{L_z}{\hbar} + \frac{1}{8}
\left(\frac{B}{B_0}\right)
^2 \frac{(x^2 + y^2)}{a_0^2} 
  - a_0 V  + \left(\frac{B}{B_0}\right) 
\frac{S_z}{\hbar}  + \frac{a_0}{R}\right],
\end{equation}
where
$
L_z = x p_y - y p_x.
$ 
Choosing  atomic units,
i.e., $m_e=\hbar=e=1$, the Hamiltonian reads
\begin{equation}
H= \frac{{p_z}^2}{2} + H_{\rm{osc}} - \frac{B}{2} L_z + B S_z - V +
\frac{1}{R}, 
\end{equation}
where
\begin{equation}
H_{\rm{osc}} = \frac{{p_x}^2+{p_y}^2}{2} + \frac{B^2}{8} ( x^2 + y^2 ).
\end{equation}

In this system of units, the magnetic field is measured in units of
$B_0=c\approx137$, the energy in units of
$\hbar \omega_0 = 1$  Hartree $= 27.2$ [eV] and the separation
$R$ in units of $a_0  = 0.53$  [\AA].
Moreover, using cylindrical coordinates (i.e., $x=\rho\cos\varphi$,
$y=\rho\sin\varphi$ and
$z=z$), 
\begin{equation}
V(z,\rho) = \frac{1}{((z-\frac{R}{2})^2 + \rho^2)^\frac{1}{2}} +
\frac{1}{((z+\frac{R}{2})^2 + \rho^2)^\frac{1}{2}}. 
\end{equation}


\section{Spectral decomposition}

In order to diagonalize $H$, we use that $[H,L_z]=[H,S_z]=[L_z,S_z]=0$ and the
spectral decompositions

\begin{equation}
L_z = \bigoplus_{m\in Z\!\!\!Z}m\Pi^{(m)} \quad \textrm{and} \quad 
S_z = \bigoplus_{s_z=\pm 1/2}s_z\Pi^{(s_z)}
\end{equation}
to get a diagonal matrix $H = \bigoplus H^{(m,s_z)} \Pi^{(m)} \otimes 
\Pi^{(s_z)}$ where
\begin{equation}
H^{(m,s_z)} = \frac{p_z^2}{2} + H_{\rm{osc}}^{(m)} -\frac{B}{2} m + B s_z - V +
\frac{1}{R},
\end{equation}
and $H_{\rm{osc}}^{(m)} = \frac{B}{2} \bigoplus_{n=0}^{\infty} (|m| +2n+1)
\Pi_n^{(m)}$. Here $\Pi_n^{(m)}$ denotes
the projector of the $n^{th}$ eigenvalue of $H_{\rm{osc}}^{(m)}$.

The ground state belongs to $s_z= -\frac{1}{2}$ in the spin sector and to $m=0$
in the $L_z$ sector,
assuming that the result of    \cite{avron} applies also to $H_2^+$. So, we
take the term
\begin{equation}
H^{(0,-\frac{1}{2})} = h^{(0,-\frac{1}{2})} + \frac{1}{R},
\end{equation}
where $h^{(0,-\frac{1}{2})} =\frac{p_z^2}{2} + H_{\rm{osc}}^{(0)} -\frac{B}{2} 
 - V $. Using the projectors
of the spectral decomposition of $H_{\rm{osc}}^{(0)}$ we can write
$h^{(0,-\frac{1}{2})}$ in a matrix form
as follows
\begin{equation}
h^{(0,-\frac{1}{2})}= \left(
\begin{array}{cc}
 \Pi_{\rm{eff}}h^{(0,-\frac{1}{2})}\Pi_{\rm{eff}}& \Pi_{\rm{eff}} h^{(0,-\frac{1}{2})}\Pi_\perp \\
 \Pi_\perp h^{(0,-\frac{1}{2})}\Pi_{\rm{eff}} & \Pi_\perp h^{(0,-\frac{1}{2})}\Pi_\perp
\end{array}
\right),
\end{equation}
where $\Pi_{\rm{eff}}=\Pi_0^{(0)}$ and $\Pi_\perp = 1-\Pi_{\rm{eff}}$, and $\Pi_0^{(0)}$ the
projector over the lowest
Landau level. The ``effective" part of our Hamiltonian, which is the physically relevant, $h_{\rm{eff}} =\Pi_{\rm{eff}}
h^{(0,-\frac{1}{2})}\Pi_{\rm{eff}}$,
can be written as 
\begin{equation}
h_{\rm{eff}} = \frac{p_z^2}{2} - V_{\rm{eff}}
\end{equation}
with
\begin{equation}
 V_{\rm{eff}}(z)=B \int_0^\infty e^{-\frac{B\rho^2}{2}} V(z,\rho) \rho d\rho,
\end{equation}
since $\Pi_{\rm{eff}}$ projects onto the wave function 
\begin{equation}
\psi_0^{(0)} (\rho)= \sqrt{\frac{B}{2\pi}} e^{-\frac{B\rho^2}{4}}.
\end{equation} 
Making the change of variable, 
$u=\frac{B\rho^2}{2}$ we get
\begin{equation}
V_{\rm{eff}}(z)= \int_0^\infty  e^{-u} \tilde V(z,u) du
\end{equation}
with
\begin{equation}
 \tilde V(z,u)= \frac{1}{\left(\left(z -\frac{R}{2} \right)^2 + 
\frac{2u}{B}\right)^{\frac{1}{2}}} + \frac{1}{\left(\left(z +\frac{R}{2}
\right)^2 + 
\frac{2u}{B}\right)^{\frac{1}{2}}}.
\end{equation}

In the sequel we present some necessary mathematical results which will appear
in full elsewhere \cite{benguria}.
Let $U_{L}$ be the unitary implementation of the scaling $z \rightarrow z/L$
and  define the scaled matrix
as  $h_L = \frac{1}{L^2}U_Lh^{(0,-\frac{1}{2})}U_L^{-1}$. Its singular part
reads 
\begin{equation}
h_{\rm{eff}}^L = \frac{p_z^2}{2} - \frac{1}{L^2}V_L,
\end{equation}
where $V_L=V_L^- + V_L^+$ and
\begin{equation}
V^{\mp}_L (z)= \int_0^\infty \frac{e^{-u}}{\left(\frac{1}{L^2}\left(z \mp
\frac{RL}{2} \right)^2 + 
\frac{2u}{B}\right)^{\frac{1}{2}}}du.
\end{equation}

If we consider the resolvents $r_L = (h_L - \xi)^{-1}$ and $r_\delta =
(h_\delta -\xi)^{-1}\otimes\Pi_{\rm{eff}} \oplus 0$
where
\begin{equation}
h_\delta= \frac{p_z^2}{2} - \delta\left(z - \frac{RL}{2}\right)- \delta\left(z
+\frac{RL}{2}\right),
\end{equation}
we can show that \cite{benguria} 
\begin{equation}
||r_L - r_{\rm{eff}}^L|| \leq \frac{C_1}{\sqrt{B}} \quad \textrm{and} \quad ||r_{\rm{eff}}^L -
r_\delta|| \leq \frac{C_2}{L}
\end{equation}
with $r_{\rm{eff}}^L = (h_{\rm{eff}}^L - \xi)^{-1}$, the resolvent of $h_{\rm eff}^L$, and
$C_1$ and $C_2$ 
are some positive constants. The appropriate scaling law for $L$ in terms of
the magnetic field is given by
$L=L(B)=2W(\frac{\sqrt{B}}{2})$, where $W(x)$ denotes the
Lambert function (for more information about this function see \cite{corless}).
This relationship between $L$ and the magnetic field $B$ is the
correct scaling law to insure the convergence of the
resolvents of the two different models which follows  from the proof of the
key theorem in  \cite{benguria}.
Such a resolvent estimate  implies, in particular,  that the ground state of
$h_L$, 
our scaled Hamiltonian, is asymptotic to $e_0$, the ground state of $h_{\delta}$,
when 
the magnetic field tends to infinity. 


\section{Perturbation theory}

In what follows, we will compute the ground state of $h_{\rm{eff}}^L$ perturbatively from
the asymptotic model $h_\delta$. 

The ground state energy, $e_0$, of $h_\delta$ is given by 

\begin{equation}
e_0 = - \frac{\alpha_0^2}{2} \quad \textrm{with} \quad \alpha_0 = 1 + 
\frac{W(2a e^{-2a})}{2a},  \label{eq22}
\end{equation}
where $a=RL/2$. Here  $\alpha_0$ is the solution of the equation
$
\alpha_0 = 1+ e^{-2a\alpha_0},
$
which can be expressed as (\ref{eq22}), in terms of the Lambert function $W$.

Since the Hamiltonians $h_L$ and $h_\delta$ are invariant under parity and
since
the ground state is even, it is convenient to work from here on  in $L^2({\bf
R}^+)$. This
being the case, the ground state of $h_\delta$ in $L^2({\bf R}^+)$ reads,  

\begin{equation}
\psi_0 (z)= \left\{
\begin{array}{cc}
 A_1 e^{-\alpha_0 z} & z > a \\
\\
 A_2  \cosh(\alpha_0 z)   & z  < a \\
\end{array} \right.,
\end{equation}
with
$
A_1 = \alpha_0 e^{2a \alpha_0} A_2 /2
$
and
$
A_2 = 2 / \sqrt{2(2a + e^{2a \alpha_0})} .
$

We set

\begin{equation}
\Delta V = h_{\rm{eff}}^L - h_\delta = \delta\left(z - a\right) - \frac{1}{L^2} V_L.
\end{equation}
To second order in perturbation theory, the ground state energy of $h_{\rm{eff}}^L$ is
given by \cite{kato}

\begin{equation}
 e_2 = e_0 + tr(P_0\Delta V P_0) - tr(P_0 \Delta V \hat{r}_\delta \Delta V P_0)
\label{eq:el-en}
\end{equation}
where $P_0$ is the projector over $\psi_0$ and $\hat{r}_\delta$ is  the
corresponding reduced resolvent.

The free kernel is given by

\begin{equation}
G_0(x,y;\xi) = \frac{1}{\sqrt{-2\xi}} \left( e^{-\sqrt{-2\xi}|x-y|} +
e^{-\sqrt{-2\xi}(x+y)} \right)
\end{equation}
and we use the notation $\partial_\xi G_0(x,y;\xi) \equiv \partial
G_0(x,y;\xi)/ \partial \xi$.
Let $G_0(x,y)$ be the free kernel evaluated at $\xi=e_0$, and correspondingly
we let 
$\partial_\xi G_0(x,y)$ be $\partial_\xi G_0(x,y;\xi)$ evaluated at $\xi=e_0$.
Using this
notation, the kernel of the reduced resolvent  at $\xi= e_0 = -\alpha_0^2/2$ 
is given by

\begin{equation}
\hat G(x,y) = G_0(x,y) 
+ \left( G_0(x,a) \quad \partial_\xi G_0(x,a) \right) 
\left(
\begin{array}{cc}
A_3 & A_4 \\
A_4 & 0
\end{array} \right) 
\left(\begin{array}{c}
 G_0(a,y) \\
\\
\partial_\xi G_0(a,y) 
\end{array} \right)
\end{equation}
with
\begin{equation}
A_3 = \frac{1}{2} \frac{\partial_\xi^2 G_0(a,a)}{\left(\partial_\xi 
G_0(a,a)\right)^2},  \qquad 
A_4 = \frac{-1}{\partial_\xi 
G_0(a,a)}.
\end{equation}

Therefore, we can compute
for each $B$ and $R$ the energy of the molecule $H_2^+$ in a magnetic field,
now
including the scaling and the repulsion energy, using the expression
\begin{equation}
E_2 (B,R) = L^2(B) e_2(B,R) +\frac{1}{R}.\label{eq29}
\end{equation}

The minimum of $E_2(B,R)$ as a function of $R$, determines the equilibrium
distance between the nuclei as well as the binding energy, which is defined as this minimum. We 
set $E_2 \equiv E_2(B,R_{eq})$, where $R_{eq}$ denotes the equilibrium
distance.

\vspace{1cm}
\section{Numerical results}

We have computed, with Mathematica \cite{mathematica}, the equilibrium distance
and
binding energy of the $H_2^+$ molecule using second order
 perturbation theory, for a wide range of magnetic fields.
Using the high level language of Mathematica, our program to compute $E_2(R)$ is simply a transcription in this
language of the mathematical formula (25), (29) and (31). Simple and double integrals
are made with the numerical integration of Mathematica using the Gauss-Kronrod method option.
Plotting $R \rightarrow E_2(R)$ we are able to locate $R_{eq}$ and finally to compute $E_2(R_{eq})$.
In tables I, II, III and IV, we compare our results for the binding
energy and internuclear separation, respectively, with those found 
by de Melo et al. \cite{melo1} (using variational techniques), Le Guillou et
al. 
\cite{guillou}, Lai et al. \cite{lai} and Heyl et al. \cite{heyl},
respectively. In figures 1 and  2 we plot the binding energy, and in figure 3
we plot the internuclear equilibrium distance, both against the 
natural scaling of these quantities, i.e., $L^2$ for the binding energy and $L$
for the internuclear distance. 
Finally, in  figure 4 we see that the product $R_{eq} L $ is monotonically
decreasing as $B$ increases.

\section{Stability of homonuclear diatomic molecules with one electron}

So far we have discussed the behavior of the ground state of $H_2^+$ for large
magnetic fields. In this
section we will consider homonuclear diatomic molecules with one electron but
with nuclear charge not necessarily one.
We are interested on the stability of these molecules as a function of $Z$ and
$B$. In particular,
we are interested in determining the highest value of $Z$ for which a diatomic
molecule of this sort may exist, as 
a function of the strength of the magnetic field. We have not try to go beyond
realistic values of $B$, i.e. $B \geq 10^{15}$. We will use the same model as
the one discussed above but this time with nuclear charge $Z$.
In order to take into account the explicit dependence of the model on $Z$ we
must change $a$ to  $  a (Z) = Z R L/2$  and $V^{\mp}_L(z)$ to 

\begin{equation}
V^{\mp}_L(z,Z) = \int_0^\infty \frac{e^{-u}}{\left(\frac{1}{L^2}\left(z \mp
\frac{ZRL}{2} \right)^2 + 
\frac{2uZ^2}{B}\right)^{\frac{1}{2}}}du
\end{equation}
and, correspondingly, the second order perturbation theory expression for the
energy of the system is given by

\begin{equation}
E_2 (B,R,Z) = Z^2\left[L^2(B) e_2(B,R,Z) +\frac{1}{R}\right]. 
\end{equation}
Here $e_2$ is given, as before, by (\ref{eq:el-en}), but with the appropriate
changes 
in $\Delta V$. 

In the case of zero magnetic field, it was shown numerically by Hogreve
\cite{hogreve}, that a
homonuclear diatomic molecule with one electron exists as long as $Z$ is below
$Z_{crit} =
1.2367$. From our calculations here (see Table V below) we see that the
magnetic field enhances the binding properties of this molecular ion. The maximum $Z$ for which a homonuclear diatomic molecule
with one electron exists, is a monotonically increasing function of $B$.
Concerning our notation in Table V,   
below $Z_c^{bs}$ we have a bound state of the molecule. Between $Z_c^{bs}$ and
$Z^{cr}$ we have only a local minimum, i.e., a resonance. Above $Z^{cr}$, we
have no minimum for the potential energy curve $R \rightarrow E_2(B,R)$,
see (\ref{eq29}), at finite values of $R$.

\section{Concluding remarks}

Applying perturbation theory to a solvable model, we have
computed the binding energy and the internuclear separation for the
$H_2^+$ molecule in the presence of a strong magnetic field.
Our results are in good agreement with those found in the literature for
a wide range of the magnetic field.
We also give the critical nuclear charge for the stability of diatomic
molecules with one electron in a strong magnetic field.
We remark that for magnetic fields about  $10^{13}$ gauss , we can find a
diatomic molecule with one electron with a nuclear charge $Z=2$, which is a new
atomic system.
Finally, we note that one advantage of perturbation theory on variational methods is to give the possibility 
of computing numerically a window which contains the exact value of the ground state energy
$E(R)$ of the effective Hamiltonian $h_{\rm{eff}}$. In other words to produce lower bounds on $E(R)$ whereas variational
methods are not good at that. Even using (21) we may compute these windows for the ground state energy of the complete
Hamiltonian $H$. We hope to come back to this question in a further work.

\section*{Acknowledgments}

This work has been supported by a CNRS(France)--Conicyt(Chile) collaborative
grant.
One of the authors (SPO) is grateful for the hospitality of CPT--Marseille
during 
the course of this work. The work of (SPO) has been supported by
FONDECYT(Chile)  project 3020050. The work of (RB) has been supported by
FONDECYT (Chile) project 102-0844.

\newpage



\begin{table}
\caption{Comparative table of data with those ones \\ 
calculated by de Melo et al.\cite{melo1} (with a $\star$).}
\begin{indented}
\item[]
\begin{tabular}{rcccc}
\br
B  (Gauss) & $R_{eq}$  (a.u.) & $-E_2$  (Hartree) & $R_{eq}^\star$  (a.u)  & $-E^\star$  (Hartree)\\
\mr
$1 \times 10^{10}$  & 1.494& 1.49 & 1.232  & 1.42 \\
$5 \times 10^{10}$  & 0.813 & 2.92 & 0.736  & 2.77 \\
$1 \times 10^{11}$  & 0.632 & 3.84 & 0.604 & 3.64  \\
$5 \times 10^{11}$  & 0.364 & 6.97 & 0.351 & 6.59 \\
$1 \times 10^{12}$  & 0.291 &  8.86 & 0.285 & 8.38 \\
$5 \times 10^{12}$  & 0.182 & 14.90 & 0.179 & 14.08 \\
$1 \times 10^{13}$  & 0.148 & 18.35 & 0.149 & 17.32 \\
$5 \times 10^{13}$  & 0.099 &  28.76 & 0.104 & 26.90 \\
$1 \times 10^{14}$  & 0.084 &  34.40 & 0.085 & 31.86 \\
\br
\end{tabular}
\end{indented}
\end{table}


\begin{table}
\caption{Comparative table of data with those ones \\  
calculated by  Guillou et al.\cite{guillou} (with a $\star$).}
\begin{indented}
\item[]
\begin{tabular}{rcccc}
\br
B  (Gauss) & $R_{eq}$  (a.u.) & $-E_2$  (Hartree) & $R_{eq}^\star$  (a.u)  & $-E^\star$  (Hartree)\\
\mr
$1.175 \times 10^{10}$  & 1.403 & 1.59 & 1.358 & 1.54  \\
$ 2.35 \times 10^{10}$ & 1.073 & 2.14 & 1.038 & 2.06   \\
$3.525 \times 10^{10}$ & 0.923 & 2.53 & 0.893 & 2.43   \\
$4.7 \times 10^{10}$ & 0.830 & 2.84 & 0.803 & 2.73 \\
$5.875 \times 10^{10}$ &0.766 & 3.11 & 0.740 & 2.98  \\
$1.175 \times 10^{11}$ & 0.596 & 4.08 & 0.578 & 3.91  \\
$ 2.35 \times 10^{11}$ & 0.467 & 5.30 & 0.455 & 5.08 \\
$ 4.7 \times 10^{11}$ & 0.371 & 6.82 & 0.362 & 6.54  \\
$ 7.05 \times 10^{11}$ & 0.325 & 7.86 & 0.318 & 7.55 \\
$1.175 \times 10^{12}$ & 0.276 & 9.36 & 0.271 & 9.01\\
$ 2.35 \times 10^{12}$ & 0.224 & 11.75 & 0.221 & 11.35 \\
$4.7 \times 10^{12} $&  0.183 &14.62 & 0.181 & 14.17  \\
\br
\end{tabular}
\end{indented}
\end{table}


\begin{table}
\caption{Comparative table of data with those ones \\
calculated by  Lai et al.\cite{lai} (with a $\star$).}
\begin{indented}
\item[]
\begin{tabular}{rcccc}
\br
B  (Gauss) & $R_{eq}$  (a.u.) & $-E_2$  (Hartree) & $R_{eq}^\star$  (a.u)  & $-E^\star$  (Hartree)\\
\mr
$1 \times 10^{11}$  & 0.632 & 3.84 & 0.61 & 3.67  \\
$5 \times 10^{11}$  & 0.364 & 6.97 & 0.35 & 6.69 \\
$1 \times 10^{12}$  & 0.291 &  8.86 & 0.280 & 8.53 \\
$2\times 10^{12} $  & 0.235 &11.16 & 0.230 & 10.78  \\
$5 \times 10^{12}$  & 0.180 & 14.90 & 0.180 & 14.46 \\
$8 \times 10^{12}$  & 0.158 & 17.18 & 0.15 & 16.71 \\
$1 \times 10^{13}$  & 0.148 & 18.35 & 0.15 & 17.88 \\
$1 \times 10^{14}$  & 0.084 &  34.40 & 0.085 & 33.83 \\
$5 \times 10^{14}$  & 0.060 & 50.60 & 0.060 & 50.07 \\
\br
\end{tabular}
\end{indented}
\end{table}


\begin{table}
\caption{Comparative table of data with those ones \\ 
calculated by  Heyl et al.\cite{heyl} (with a $\star$).}
\begin{indented}
\item[]
\begin{tabular}{rcccc}
\br
B  (Gauss) & $R_{eq}$  (a.u.) & $-E_2$  (Hartree) & $R_{eq}^\star$  (a.u)  & $-E^\star$  (Hartree)\\
\mr
$9.4  \times 10^{12}$ &0.151 & 18.02 & - & 17.52 \\
$2.35 \times 10^{13}$ & 0.119 & 23.43 & - & 22.89  \\
$4.7 \times 10^{13}$ & 0.100 & 28.29 & - & 27.69 \\
$9.4 \times 10^{13}$  & 0.085 & 33.87 & -  & 33.28 \\
$2.35 \times 10^{14}$ &  0.070 & 42.44 & - & 41.73 \\
$4.7 \times 10^{14}$ &0.061 & 49.89 & - & 49.14 \\
\br
\end{tabular}
\end{indented}
\end{table}


\begin{table}
\caption{Data of the stability study of the system}
\begin{indented}
\item[]
\begin{tabular}{ccccccc}
\br
   &  Z=1    &&  Z=1.2  && Z=1.4    &\\
\mr
B  (Gauss) & $R_{eq}$  (a.u.)& $-E_2 $ (Hartree)  & $R_{eq}$  (a.u.) & $-E_2$  (Hartree) &
$R_{eq}$ (a.u.) & $-E_2 $  (Hartree) \\
\mr
$1\times 10^{10}$&1.494 &1.49 &1.718 &1.71&  & \\
$1\times 10^{11}$ &  0.632 & 3.84  & 0.712 & 4.51&0.810&5.19\\
$1\times 10^{12}$ &  0.291 &  8.86  &0.318 &10.58 &0.353&12.27\\
$1\times 10^{13}$ &  0.148 & 18.36  &0.157 &22.23&0168&25.98\\
$1\times 10^{14}$ &  0.084 & 34.40  &0.087 &42.33&0.090&50.05\\
\br
\end{tabular}
\end{indented}
\end{table}

\begin{table}
\begin{indented}
\item[]
\begin{tabular}{ccccccc}
\br
  &Z=1.6 &&  Z=1.8  &&  Z=2 \\
\mr
B  (Gauss) & $R_{eq}$   (a.u.)& $-E_2$  (Hartree) &$R_{eq}$   (a.u.)& $-E_2$  (Hartree) &$R_{eq}$   (a.u.)& $-E_2$  (Hartree)\\
\mr
$1\times 10^{12}$ & 0.400 & 13.99&0.445 &15.78 && \\
$1\times 10^{13}$ & 0.183 & 29.67&0.202 &33.39  &  0.227 & 37.26  \\
$1\times 10^{14}$ & 0.095 & 57.55&0.101 &64.88  &  0.108 & 72.14   \\
\br
\end{tabular}
\end{indented}
\end{table}

\begin{table}
\begin{indented}
\item[]
\begin{tabular}{ccccc}
\br
  &  Z=2.2  && Z=2.4      \\
\hline
\hline
B  (Gauss)  & $R_{eq}$ (a.u.) & $-E_2$  (Hartree)& $R_{eq}$  (a.u.)& $-E_2$  (Hartree)\\
\mr        
$1\times 10^{14}$ & 0.118 & 79.44&0.131& 86.98     \\
\br
\end{tabular}
\end{indented}
\end{table}


\begin{table}
\caption{Upper bounds to  the critical nuclei charge of stability of the molecule given by our model.}
\begin{indented}
\item[] \begin{tabular}{ccc}
\br
B  (Gauss) & $Z_{c}^{bs}$  & $Z_{}^{cr}$\\
\mr
$1\times 10^{10}$ &      $ < 1.32 $ &$ < 1.58$\\
$1\times 10^{11}$ &      $ < 1.55$  &$ < 2.05$\\
$1\times 10^{12}$ &      $ < 1.80$  &$ < 2.60$\\
$1\times 10^{13}$ &      $ < 2.10$  &$ < 3.10$\\
$1\times 10^{14}$ &      $ < 2.43$  &$ < 3.50$\\
\br
\end{tabular}
\end{indented}
\end{table}


\begin{figure}
\begin{center}
\epsfxsize= 13cm
\epsfbox{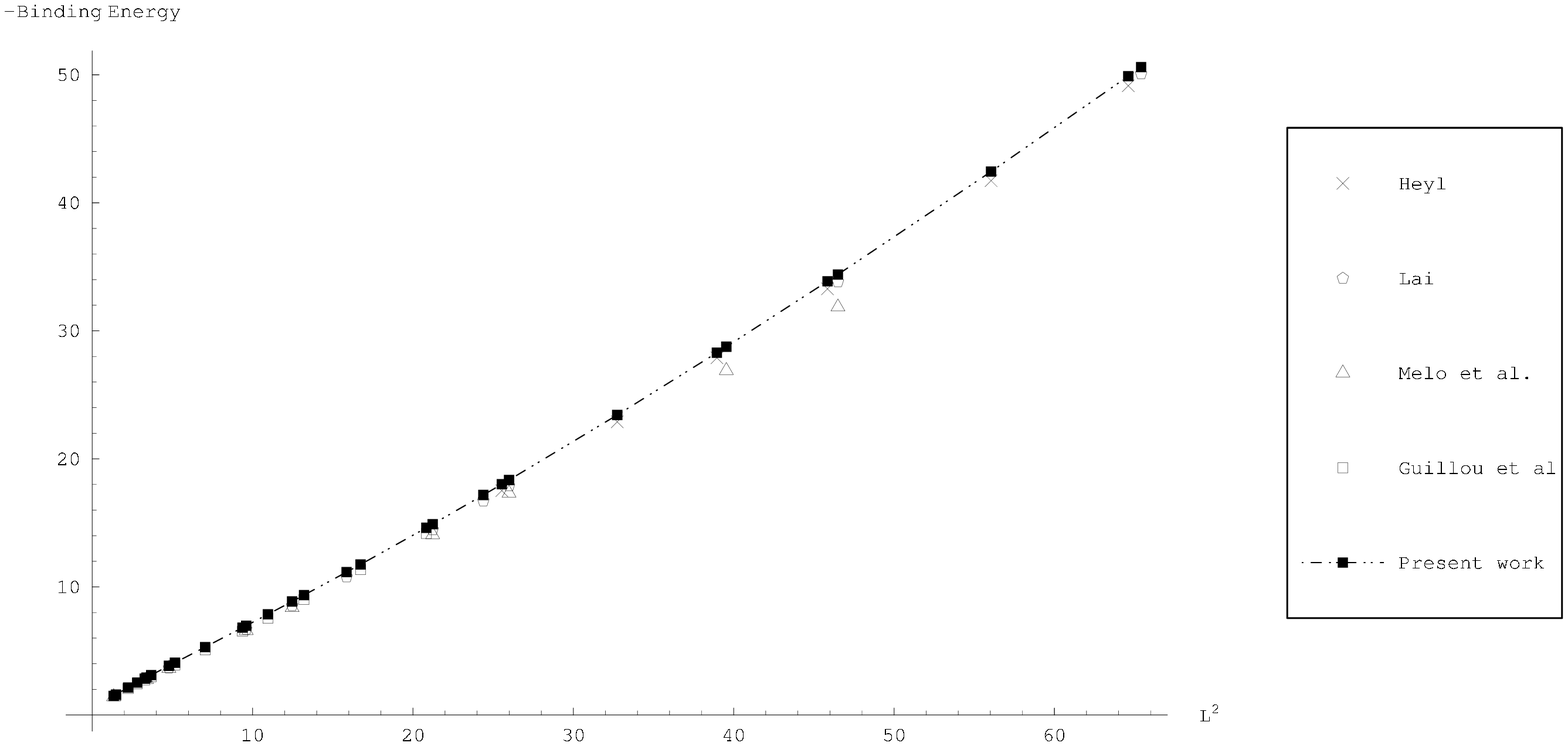}
\end{center}
\caption{Comparison of the binding energies for all magnetic field values. The binding energy is plotted in atomics units.
The adimensional scaling parameter $L$ is calculated as $L=2W(\frac{1}{2}\sqrt{\frac{B}{B_0}})$ and therefore is independent
of system of units chosen for the magnetic field.}
\end{figure}

\begin{figure}
\begin{center}
\epsfxsize= 13cm
\epsfbox{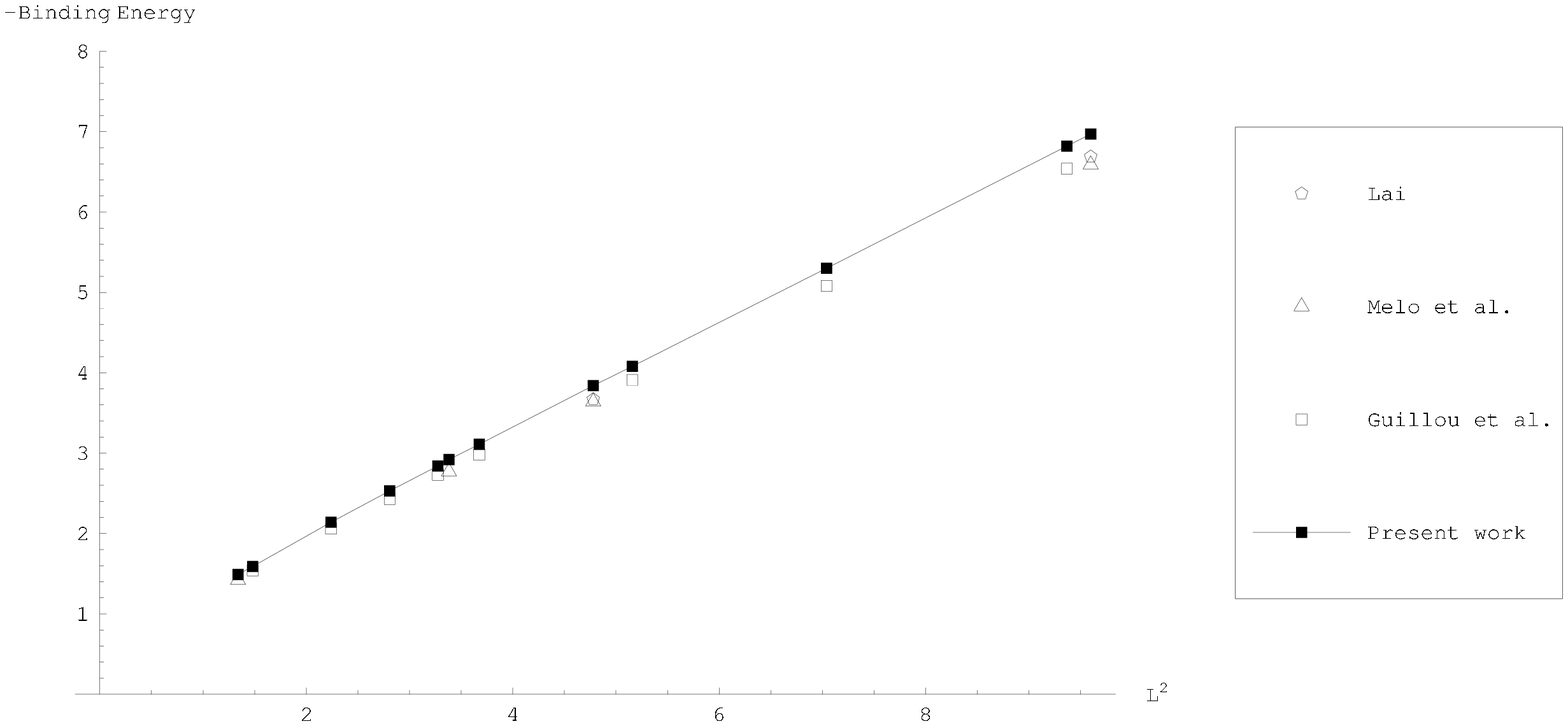}
\end{center}
\caption{Detail of the  figure 1}
\end{figure}


\begin{figure}
\begin{center}
\epsfxsize= 13cm
\epsfbox{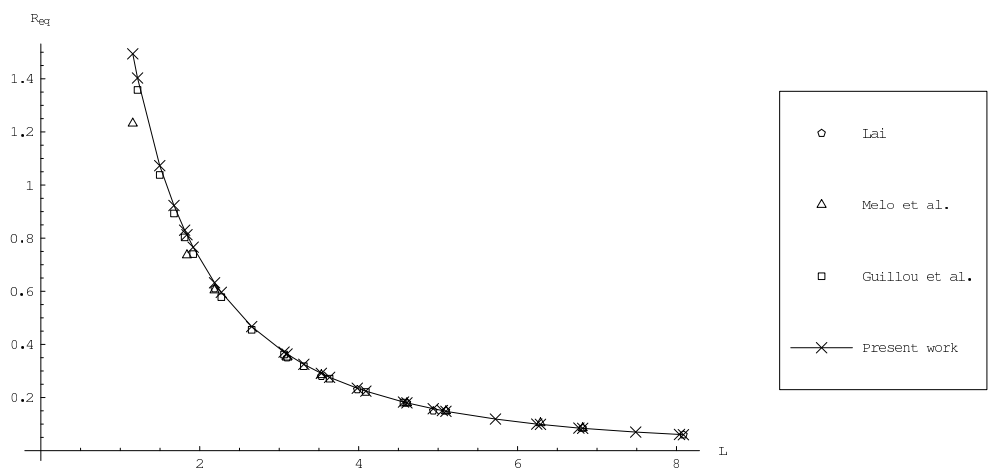}
\end{center}
\caption{Comparison of the equilibrium internuclear distance, which is  plotted in atomic units.}
\end{figure}

\begin{figure}
\begin{center}
\epsfxsize= 13cm
\epsfbox{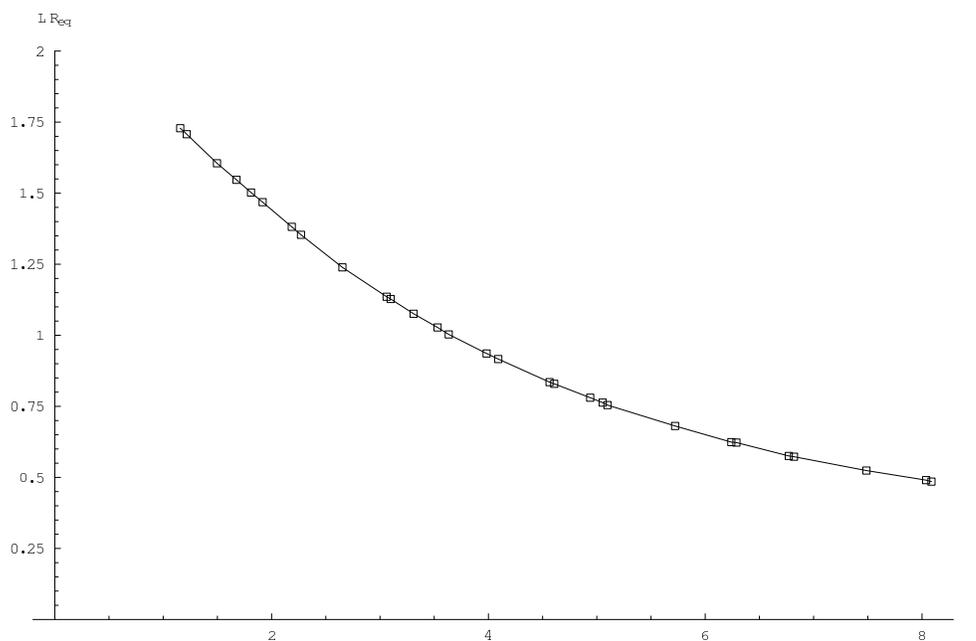}
\end{center}
\caption{Behaviour of the product of  $ \textrm{R}_{eq}  $ L}
\end{figure}

\end{document}